\documentclass[twocolumn]{revtex4}

\usepackage{latexsym}
\usepackage{amssymb}
\usepackage{epsfig}
\usepackage{amsmath}
\usepackage{float}
\usepackage{color}
\usepackage{enumitem}
\usepackage{placeins}

\newcommand{\adotoa}{\ensuremath{{\cal H}}}

\newcommand{\grad}{\ensuremath{\vec{\nabla}}}

\newcommand{\Pb}{\ensuremath{\bar{P}}}
\newcommand{\rhob}{\ensuremath{\bar{\rho}}}

\newcommand{\phib}{\ensuremath{\bar{\phi}}}
\newcommand{\Zb}{\ensuremath{\bar{Z}}}

\newcommand{\rhop}{\ensuremath{\rho_{\phi}}}

\newcommand{\thetap}{\ensuremath{\theta_{\phi}}}

\newcommand{\be}{\begin{equation}}
\newcommand{\ee}{\end{equation}}
\newcommand{\bea}{\begin{eqnarray}}
\newcommand{\eea}{\end{eqnarray}}

\newcommand{\fref}[1]{figure~\ref{#1}}
\newcommand{\sref}[1]{section~\ref{#1}}

\newcommand{\MontePython}{\textsc{MontePython}}
\newcommand{\CLASS}{\textsc{class}}

\begin{document}

\title{ Reconciling CMB and structure growth measurements with dark energy interactions } 

\author{
Alkistis Pourtsidou\footnote{E-mail: alkistis.pourtsidou@port.ac.uk} and
Thomas Tram\footnote{E-mail: thomas.tram@port.ac.uk}
}

\affiliation{
Institute of Cosmology \& Gravitation, University of Portsmouth, Dennis Sciama Building, Burnaby Road, Portsmouth, PO1 3FX, United Kingdom
}

\begin{abstract}
We study a coupled quintessence model with pure momentum exchange and present the effects of such an interaction on the Cosmic Microwave Background (CMB) and matter power spectrum. For a wide range of negative values of the coupling parameter $\beta$ structure growth is suppressed and the model can reconcile the tension between Cosmic Microwave Background observations and structure growth inferred from cluster counts. We find that this model is as good as $\Lambda$CDM for CMB and baryon acoustic oscillation (BAO) data, while the addition of cluster data makes the model strongly preferred, improving the best-fit $\chi^2$-value by more than $16$.
\end{abstract}

\maketitle

\section{Introduction}

During the last two decades observational cosmology has entered an era of unprecedented precision. Cosmic microwave background (CMB) measurements~\cite{Bennett:2012zja, Ade:2013zuv}, baryon acoustic oscillations (BAO)~\cite{Lampeitl:2009jq} and observations of Type Ia Supernovae~\cite{Riess:2009pu} have shown very good agreement with the predictions of the standard cosmological model ($\Lambda$CDM) consisting of dark energy in the form of a cosmological constant $\Lambda$ and cold dark matter (CDM). However this agreement is not perfect: the Planck CMB data~\cite{Ade:2013zuv} are in tension with low redshift data such as cluster counts~\cite{Ade:2013lmv}, redshift space distortions (RSD)~\cite{Samushia:2012iq, Macaulay:2013swa}, weak lensing data~\cite{Heymans:2012gg} and local measurements of the Hubble constant, $H_0$~\cite{Riess:2011yx,Riess:2016jrr}. More specifically, the low redshift probes point towards a lower rate of structure growth (equivalently, a lower $\sigma_8$) than the Planck  results for the base $\Lambda$CDM would prefer.

The most significant tensions are the ones coming from the cluster and weak lensing data.
A possible explanation for these tensions is that there are systematic effects that have not been accounted for, such as systematics that affect the determination of the mass bias in the cluster case, and small scales effects in the weak lensing case. Another possibility is that $\Lambda$CDM is not the correct model describing the evolution of the Universe.  Finding a different model which gives a better (or, at the very least, equally good) fit to the data has proven very difficult. However, the motivation for exploring alternatives is strong because of the fundamental problems that plague $\Lambda$CDM, namely the \emph{fine-tuning} and \emph{coincidence} problems. These problems are associated with the cosmological constant and they have led to a plethora of alternative scenarios. One popular example is quintessence~\cite{Wetterich88, Ratra:1987rm, Wetterich:1994bg}, in which an evolving scalar field plays the role of dark energy. A different viewpoint suggests that General Relativity is modified on cosmological scales and this modification is responsible for the accelerated expansion of the Universe (see~\cite{Copeland:2006wr, CliftonEtal2011} and references therein).

The nature of the two constituents of the dark sector is currently unknown and the fact that they are considered uncoupled in $\Lambda$CDM is just an assumption of the model. Let us now consider a non-gravitational coupling between cold dark matter (CDM) and dark energy (DE). The energy momentum tensors for CDM and DE are then no longer separately conserved but instead we find
\be
\nabla_\mu T^{{\rm (CDM)}\mu}_{\phantom{{\rm (CDM)}\mu} \nu} = J_\nu = -\nabla_\mu T^{{\rm (DE)}\mu}_{\phantom{{\rm (DE)}\mu} \nu},
\ee where the coupling current $J_\nu$ represents the energy and momentum exchange between dark energy and dark matter. Note that we assume that the standard model is not coupled to the dark sector --- an assumption which well justified by observations that strongly constrain such couplings~\cite{Carroll:1998zi}. 

Traditionally DE is modelled by a quintessence field $\phi$, so we are going to set ${\rm{DE}}\equiv \phi$ from now on, and ${\rm{cdm}}\equiv c$.
In the background, this means that the energy conservation relations become
\begin{align} \nonumber
\dot{\rhob}_c+3{\cal H}\rhob_c &= Q \\
\dot{\rhob}_\phi+3{\cal H}\rhob_\phi (1+w_\phi) &= -Q, 
\end{align} where ${\cal H}=\dot{a}/a$ is the conformal Hubble rate with scale factor $a$ and a dot denotes the derivative with respect to conformal time $\tau$. $\rhob_c, \rhob_\phi$ are the energy densities of cold dark matter and dark energy (with the bar denoting background quantities) and $Q \equiv \bar{J}_0$ is the background energy transfer.

The form of $Q$ is usually chosen phenomenologically, and one of the most widely used forms is 
$
Q = \alpha_0 \dot{\phib} \rhob_c
$~\cite{Amendola:1999er}, where $\alpha_0$ is a constant parameter which determines the strength of the interaction. 
This case has been extensively studied in the literature, see e.g.~\cite{Amendola:1999er, Xia:2009zzb, Xia:2013nua,Pourtsidou:2013nha} and references therein. Other widely used forms take $Q$ to be proportional to the energy densities of cdm and/or DE, i.e. $Q \propto \Gamma \rho $ with $\Gamma$ a constant interaction rate, or $Q \propto {\cal H} \rho$, see e.g.~\cite{CalderaCabral:2008bx, CalderaCabral:2009ja,Valiviita:2009nu, Jackson:2009mz,Clemson:2011an,Gavela:2009cy,Yang:2014gza,Salvatelli:2013wra, yang:2014vza, Murgia:2016ccp} and references therein. 
Interest in interacting dark energy models has grown as they are arguably better motivated than the uncoupled $\Lambda$CDM concordance model, and they can possibly lift the tensions present in the available data. For example, an interaction between vacuum energy and dark matter was shown to be able to remove some of the tensions~\cite{Salvatelli:2014zta, Wang:2015wga}. 
Very recently there have also been important theoretical developments in the field. 
In \cite{DAmico:2016aa}, the authors formulate a quantum field theory of interacting dark matter - dark energy and show that the quantum corrections coming from the usual assumption for the form of the interaction (i.e. that dark matter is made up of ``heavy" particles whose mass depend on the dark energy field value) are huge and severely constrain the allowed couplings and the nature of the dark sector (see also \cite{Marsh:2016aa}).

However, in the majority of cases there is no Lagrangian description of the model, and an ad-hoc expression for the coupling $Q$ is just added at the level of the equations. In~\cite{Pourtsidou:2013nha} the authors used the pull-back formalism for fluids to generalise the fluid action and involve couplings between the scalar field playing the role of dark energy, and dark matter. Their construction led to three distinct \emph{families} of coupled models. The first two (Types 1 and 2) involve both energy and momentum transfer between dark matter and dark energy. The commonly used coupled quintessence model~\cite{Amendola:1999er} was shown to be a sub-class of Type 1~\cite{Pourtsidou:2013nha, Skordis:2015yra}. The third family of models (Type 3) is a pure momentum transfer theory with $Q=0$. 

In this work we investigate a specific Type 3 model that was first presented in~\cite{Pourtsidou:2013nha}. We implement this model in the Einstein-Boltzmann solver \CLASS{}~\cite{Blas:2011rf}, perform a global fitting of cosmological parameters using the Markov chain Monte Carlo (MCMC) code~\MontePython{}~\cite{Audren:2012wb}, and compare our findings to $\Lambda$CDM. The structure of the paper is as follows: In Section~\ref{sec:T3model} we present the specific Type 3 model we are going to study and state the background and linear perturbation cosmological equations which were derived in detail in~\cite{Pourtsidou:2013nha}. We also demonstrate the coupling's effects on the CMB and linear matter power spectra using \CLASS{}. In Section~\ref{sec:results} we describe the datasets and priors we use and then present the results of our MCMC analysis. We concentrate on the comparison of the Type 3 model with $\Lambda$CDM and the effects on the $(\sigma_8, H_0)$ parameters. We present the best fit cosmological parameters for $\Lambda$CDM and Type 3 and compare their $\chi^2$ values. We conclude in Section~\ref{sec:conclusions}.

\section{The pure momentum transfer model}
\label{sec:T3model}

As we have already stated, the model we are going to investigate belongs to the Type 3 class of theories constructed in~\cite{Pourtsidou:2013nha}. The distinctive characteristic of this class is that there is only momentum transfer between the two components of the dark sector. No coupling appears at the background level, regarding the fluid equations --- they remain the same as in the uncoupled case. Furthermore, the energy-conservation equation remains uncoupled even at the linear level. Hence, the theory provides for a pure momentum-transfer coupling at the level of linear perturbations. 

Type 3 theories are classified via the Lagrangian~\cite{Pourtsidou:2013nha}
\begin{equation}
L(n,Y,Z,\phi) = F(Y,Z,\phi) + f(n),
\end{equation} where $n$ is the fluid number density, $Y=\frac{1}{2}\nabla_\mu \phi \nabla^\mu \phi$ is used to construct a kinetic term for $\phi$, and $Z=u^\mu \nabla_\mu \phi$ plays the role of a direct coupling of the fluid velocity $u^\mu$ to the gradient of the scalar field \footnote{ 
Note that the fluid velocity $u^\mu$ is uniquely defined in terms of the fluid number density $n$ and the dual number density (see \cite{Pourtsidou:2013nha} for details).}.
Considering a coupled quintessence function of the form 
\be
F=Y+V(\phi)+h(Z),
\ee with $V(\phi)$ the quintessence potential, we have the freedom to choose the coupling function $h(Z)$ in order to construct specific models belonging to the same class. Following~\cite{Pourtsidou:2013nha}, we are going to concentrate on the sub-case with 
\be
h(Z) = \beta Z^2,
\ee where $\beta$, the coupling parameter, is taken to be constant. 
Defining $\tilde{g}^{\mu \nu}=g^{\mu \nu}+2\beta u^\mu u^\nu$, we can write the action for the scalar field $\phi$ as \cite{Pourtsidou:2013nha}
\bea
 \nonumber
S_{\phi}&=&
-\int d^4x \sqrt{-g}\left[
   \frac{1}{2}\tilde{g}^{\mu \nu} \phi_\mu \phi_\nu
+ V(\phi)
\right]
\\
&\rightarrow & 
\int dt \, d^3x \, a^3 \left[
    \frac{1}{2}(1 - 2 \beta)  \dot{\phi}^2
 - \frac{1}{2} |\grad\phi|^2
- V(\phi)
\right] \;\;\;
\label{eq:action}
\eea
where the arrow denotes working in a frame where the CDM $3$-velocity is zero.
Hence, the model is physically acceptable for $\beta < \frac{1}{2}$. For $\beta \rightarrow 1/2$
we have a strong coupling problem, while for $\beta>1/2$ there is a ghost in the theory since the kinetic term becomes negative~\cite{Pourtsidou:2013nha}.

\subsection{Background Evolution}

We assume a Universe described by a flat Friedmann-Lema\^itre-Robertson-Walker (FLRW) metric
\be
ds^2=a^2(\tau)(-d\tau^2+ dx_idx^i),
\ee where $a(\tau)$ satisfies the Friedmann equation
\be
{\cal H}^2 =\frac{8\pi G}{3}a^2\rhob_{\rm tot} 
= \frac{8\pi G}{3}a^2(\rhob_\gamma + \rhob_b + \rhob_c + \rhob_\phi).
\ee Here $\rhob_{\rm tot}$ is the total background energy density of all species and the subscripts $\gamma, b$ denote radiation and baryons, respectively. 

The background energy density and pressure for quintessence are~\cite{Pourtsidou:2013nha}
\bea
\rhob_{\phi}&=&\left(\frac{1}{2}-\beta\right)\frac{\dot{\phib}^2}{a^2}+V(\phi), \\ 
\Pb_{\phi}&=&\left(\frac{1}{2}-\beta\right)\frac{\dot{\phib}^2}{a^2}-V(\phi),
\eea
and the energy conservation equations are the same as in uncoupled quintessence:
\bea
\label{eq:drhophiType3}
\dot{\rhob}_\phi &+&  3{\cal H}(\rhob_\phi+\Pb_\phi)=0, \\
\dot{\rhob}_c  &+&  3{\cal H}\rhob_c=0, 
\eea i.e. $Q=0$ for these models.
Hence the cold dark matter density obeys the usual scaling relation:
\be
\rhob_c = \rho_{c,0} a^{-3}.
\ee
In contrast to the usual coupled dark energy models which exhibit background energy exchange, the evolution of the CDM and quintessence energy densities
\be
\Omega_c \equiv \frac{\rhob_c}{\rhob_{\rm tot}}, \;\;\; \Omega_\phi \equiv \frac{\rhob_\phi}{\rhob_{\rm tot}} 
\ee are unchanged in Type 3 models. The background Klein-Gordon equation is given by
\be
\ddot{\phib}+2{\cal H}\dot{\phib}+
\left(\frac{1}{1-2\beta}\right)a^2\frac{dV}{d\phi}=0,
\ee
and the speed of sound is $c^2_s=\frac{1-2\beta}{1-2\beta}=1$~\cite{Pourtsidou:2013nha}.

\subsection{Linear Perturbations}

In order to study the observational effects of the coupled models on the Cosmic Microwave Background and Large Scale Structure (LSS), we need to consider linear perturbations around the FLRW background. Choosing the synchronous gauge the metric is
\begin{eqnarray}
ds^2 &=& -a^2d\tau^2 + a^2 \left[ (1 + \frac{1}{3}h)\gamma_{ij} + D_{ij}\nu\right] dx^i dx^j,
\ \ 
\label{perturbed_metric}
\end{eqnarray}
 where $\tau$ is the conformal time, $\grad_k$ is the covariant derivative associated with $\gamma_{ij}$, i.e. $\grad_k \gamma_{ij} = 0$
and $D_{ij}$ is the traceless derivative operator $D_{ij} = \grad_i \grad_j -\frac{1}{3} \grad^2 \gamma_{ij}$.
The unit-timelike vector field $u^\mu$ is perturbed as
\begin{eqnarray}
u_\mu &=& a( 1, \grad_i \theta).
\label{u_mu_theta}
\end{eqnarray}
We denote the field perturbation $\delta \phi$ by $\varphi$, the $\phi$-derivative of the potential by $V_\phi \equiv dV/d\phi$ and we also have $\Zb \equiv -\dot{\phib}/a$. In this notation the perturbed scalar field energy density and pressure are given by~\cite{Pourtsidou:2013nha}
\begin{align}
\delta \rhop  = -\frac{1}{a}\Zb (1  - 2\beta) \dot{\varphi}
  +  V_{\phi} \varphi, \\
\delta P_\phi = - \frac{1}{a} \Zb \left(1 - 2\beta \right)\dot{\varphi} - V_{\phi}\varphi,
\end{align}
while the velocity divergence of the scalar field is
\begin{equation}
\thetap =\frac{1}{(2\beta -1)\Zb}\left(\frac{1}{a}\varphi+ 2\beta \Zb \theta_c\right).
\label{eq:thetaphiType3}
\end{equation}
This is one other significant difference from other types of coupling, namely that $\thetap$ depends also on the CDM velocity divergence $\theta_c$.

The linearised scalar field equation is~\cite{Pourtsidou:2013nha}
\begin{eqnarray}
&&
   (1  - 2\beta) (\ddot{\varphi}    + 2 \adotoa \dot{\varphi})
+  \left(k^2 + a^2 V_{\phi\phi} \right) \varphi
\nonumber 
\\
&&
\ \ \ \
+ \frac{1}{2}  \dot{\phib} (1 - 2\beta) \dot{h}  
- 2\beta \dot{\phib}k^2  \theta_c  
= 0,
\label{eq:pertKG}
\end{eqnarray}
while the density contrast $\delta_c \equiv \delta \rho_c / \rhob_c$ obeys the standard evolution equation
\be
\label{eq:deltac}
\dot{\delta}_c = -k^2 \theta_c - \frac{1}{2}\dot{h}.
\ee 
The momentum-transfer equation depends on the coupling and is given by
\begin{equation}
 \dot{\theta}_c = - \adotoa  \theta_c + 
 \frac{( 6 \adotoa \beta \Zb + 2\beta \dot{\Zb} )  \varphi +  2\beta \Zb  \dot{\varphi}}{a \left(\rhob_c  - 2 \beta \Zb^2 \right)} .
 \label{dP_type3_pert}
\end{equation}

We implemented the above equations in \CLASS{} in order to compute the CMB temperature and matter power spectra. At this stage we will fix our quintessence potential $V(\phi)$ to be the widely used single exponential form (1EXP)
\be
V(\phi)=V_0 e^{-\lambda \phi}.
\ee In order to demonstrate the effect of the coupling before we perform an MCMC analysis, we will first compare the uncoupled $\beta = 0$ case with 
the coupled model under consideration keeping the parameter $\lambda$ of the potential fixed ($\lambda=1.22$), while the potential normalisation $V_0$ is varied automatically by \CLASS{} is order to match a fixed $\Omega_\phi$ today. Our initial conditions for the quintessence field are $\phi_i=10^{-4}, \dot{\phi}_i=0$. Note, however, that in this case and using the 1EXP potential, the same cosmological evolution is expected for a wide range of initial conditions.

Before we present our results, we should briefly discuss the range of $\beta$ values we are going to consider. As we have already mentioned, it is clear from the Lagrangian in Equation~(\ref{eq:action}) that we are not allowed to consider the case $\beta > 1/2$ due to the strong coupling and ghost pathologies. We are free to consider any negative value $\beta$, but the fact that $\beta$ is a dimensionless coupling parameter in the Lagrangian suggests that its magnitude should be small. We will therefore initially choose a prior for our negative $\beta$ values such that $-0.5 \leq \beta \leq 0$ and call this model T3. However, as we shall see later the data allow for the model to span a much wider range of negative $\beta$, so in this case we will use a log prior $-3 \leq {\rm Log}_{10}(-\beta) \leq 7$ and denote this (phenomenological) Type 3 model $\text{T3}_\text{ph}$.  

\begin{figure*}[tb]
\includegraphics[width=0.495\linewidth]{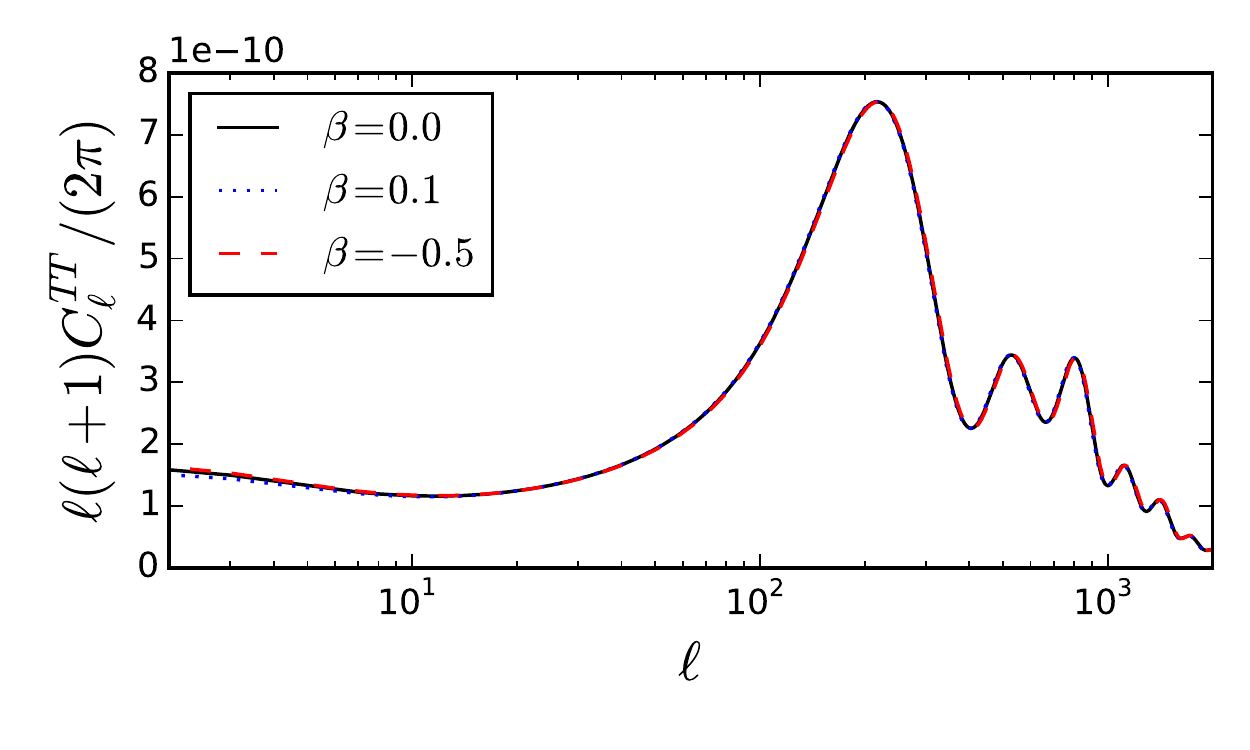}
\hfill
\includegraphics[width=0.495\linewidth]{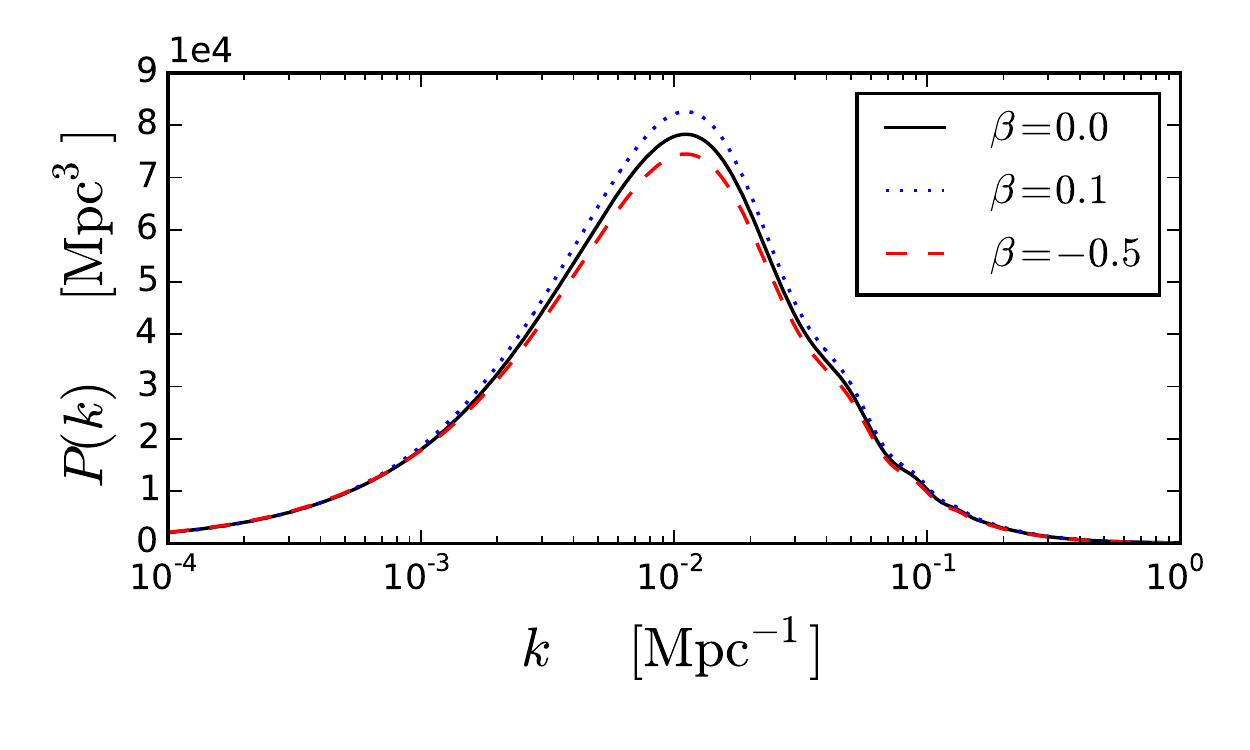}
\caption{Comparison of the CMB TT power spectra (left panel) and the linear (total) matter power spectra $P(k)$ at $z=0$ (right panel).  Black solid lines denote the uncoupled quintessence model, blue dotted lines denote the coupled model with positive coupling parameter $\beta=0.1$, and dashed red lines denote the coupled model with negative coupling parameter $\beta=-0.5$.}
\label{fig:ClTT}
\end{figure*}

In Fig.~\ref{fig:ClTT} we show the CMB temperature spectra and the matter power spectra for the uncoupled case $\beta=0$, for a positive coupling parameter $\beta=0.1$, and for $\beta=-0.5$. We see that the effect of the coupling in the CMB temperature power spectrum is very small. There is no visible effect on the small scale amplitude and no shift of the location of the peaks, like in usual coupled quintessence models (\cite{Amendola:1999er, Xia:2009zzb, Xia:2013nua}). The only visible effect is an integrated Sachs-Wolfe (ISW) effect on large scales, but it remains small in contrast to coupled quintessence models with background energy exchange. This is because the background CDM energy density remains uncoupled in our model. 
However, the effects on the matter power spectrum $P(k)$ are significant. For the positive coupling case there is enhanced growth at small scales (and, consequently, a larger value of $\sigma_8$ is obtained), while for the negative coupling T3 case the growth is suppressed and the associated $\sigma_8$ is smaller. Since the positive coupling model results in enhanced growth, it aggravates the tension between the Planck CMB data and low redshift observations. We will therefore exclude the $0<\beta<1/2$ branch from our MCMC analysis.

\begin{figure*}[tb]
\centering
\includegraphics[width=0.495\linewidth]{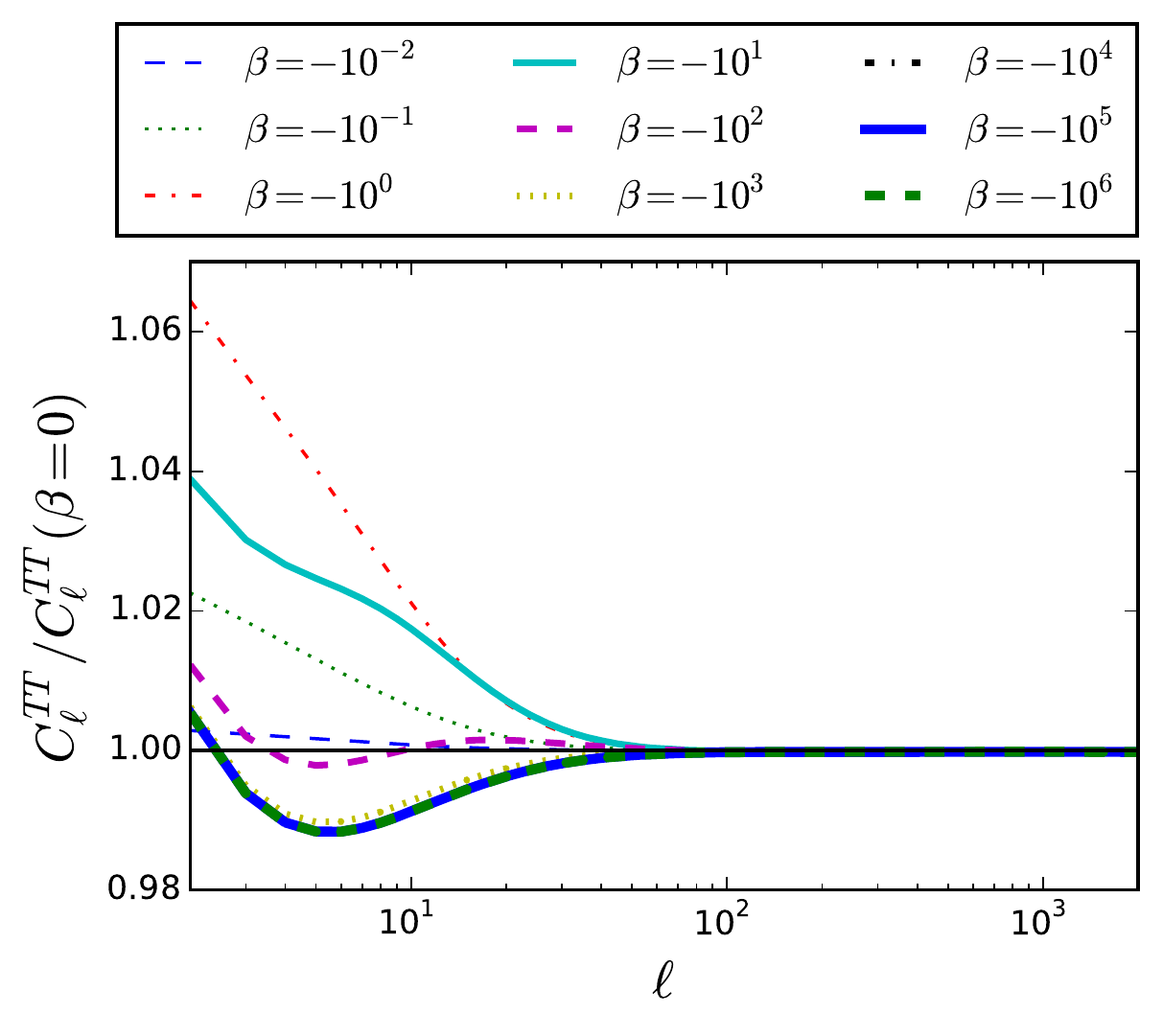}
\hfill
\includegraphics[width=0.495\linewidth]{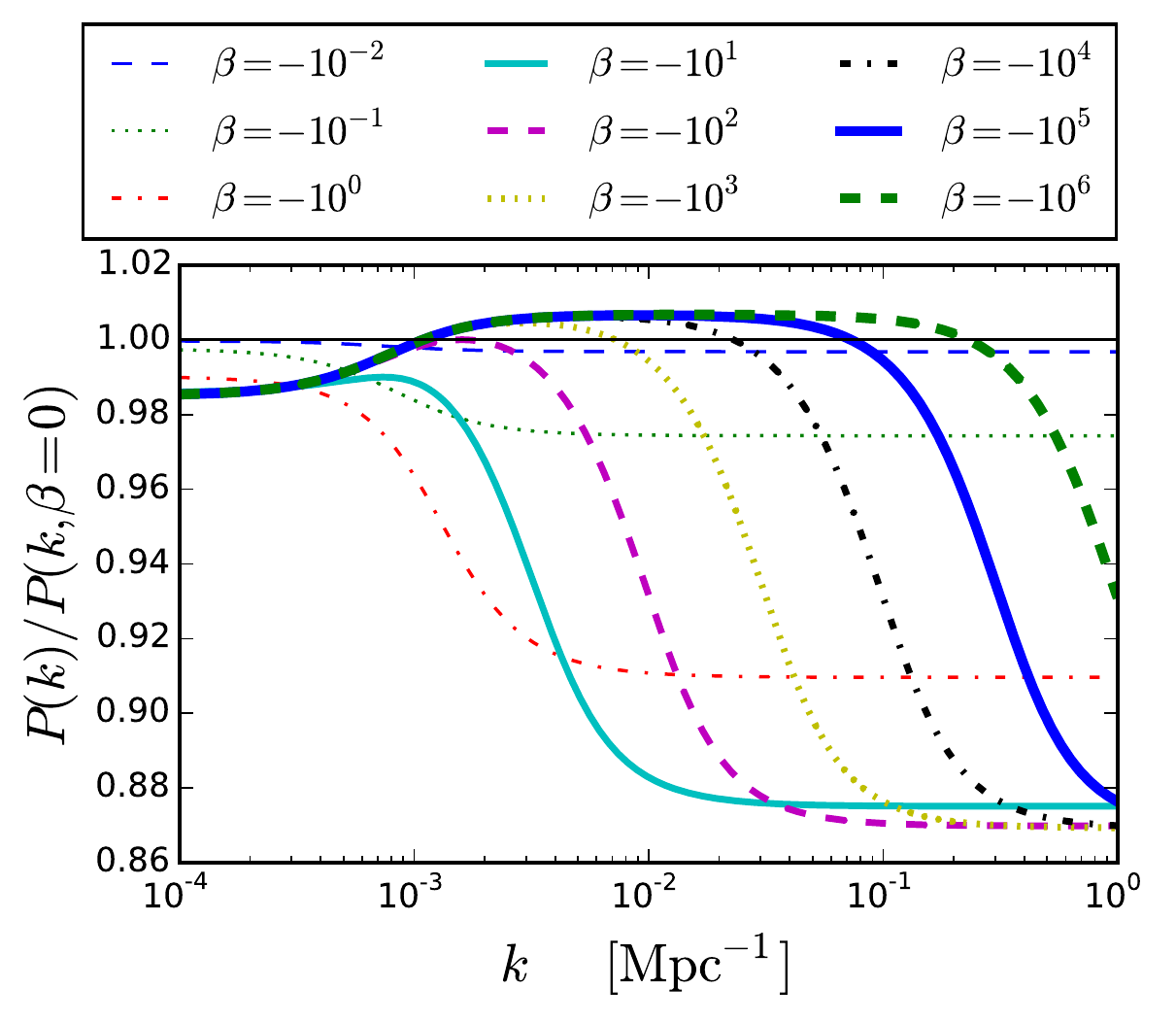}
\caption{Comparison of the CMB TT power spectra (left panel) and the linear (total) matter power spectra at $z=0$ (right panel) for a wide range of negative $\beta$-values. In both cases we show the ratio between the $\text{T3}_\text{ph}$ model and the predictions of the uncoupled quintessence model.}
\label{fig:ClTT_T3ph}
\end{figure*}

We will now show the comparison between the $\text{T3}_\text{ph}$ model - which spans a wide range of negative $\beta$ values - and the uncoupled quintessence case. In order to highlight the effect of $\beta$ we fix the sound speed at recombination $\theta_s$ and the physical energy densities of CDM and baryons, $\omega_{c,b}=\Omega_{c,b}h^2$. In Fig.~\ref{fig:ClTT_T3ph} we plot the CMB TT power spectrum and the matter power spectrum divided by their uncoupled counterparts. Similar to the T3 model, we see that the effect of the coupling on the TT power spectrum only shows up on very large scales. This is due to the late-time ISW effect which changes because of the effect of the coupling on the evolution of the matter density perturbations. We also notice that the effect changes direction (from enhancement to suppression with respect to the uncoupled case) around $\beta \simeq-10^2$. 
In the matter power spectrum we also see an interesting feature, namely that for $\beta<-10^2$ there is a range of $k$-values where the matter power spectrum is actually~\emph{enhanced} compared to the uncoupled case. This gives $\sigma_8(\beta)>\sigma_8(\beta=0)$ for $\beta\lesssim-10^4$. Note that we are comparing models that reproduce the same TT-power spectrum at small scales according to the left panel of Fig.~\ref{fig:ClTT_T3ph}, and these models have slightly different values of $H_0$.

\begin{figure}[tb]
\centering
\includegraphics[width=\columnwidth]{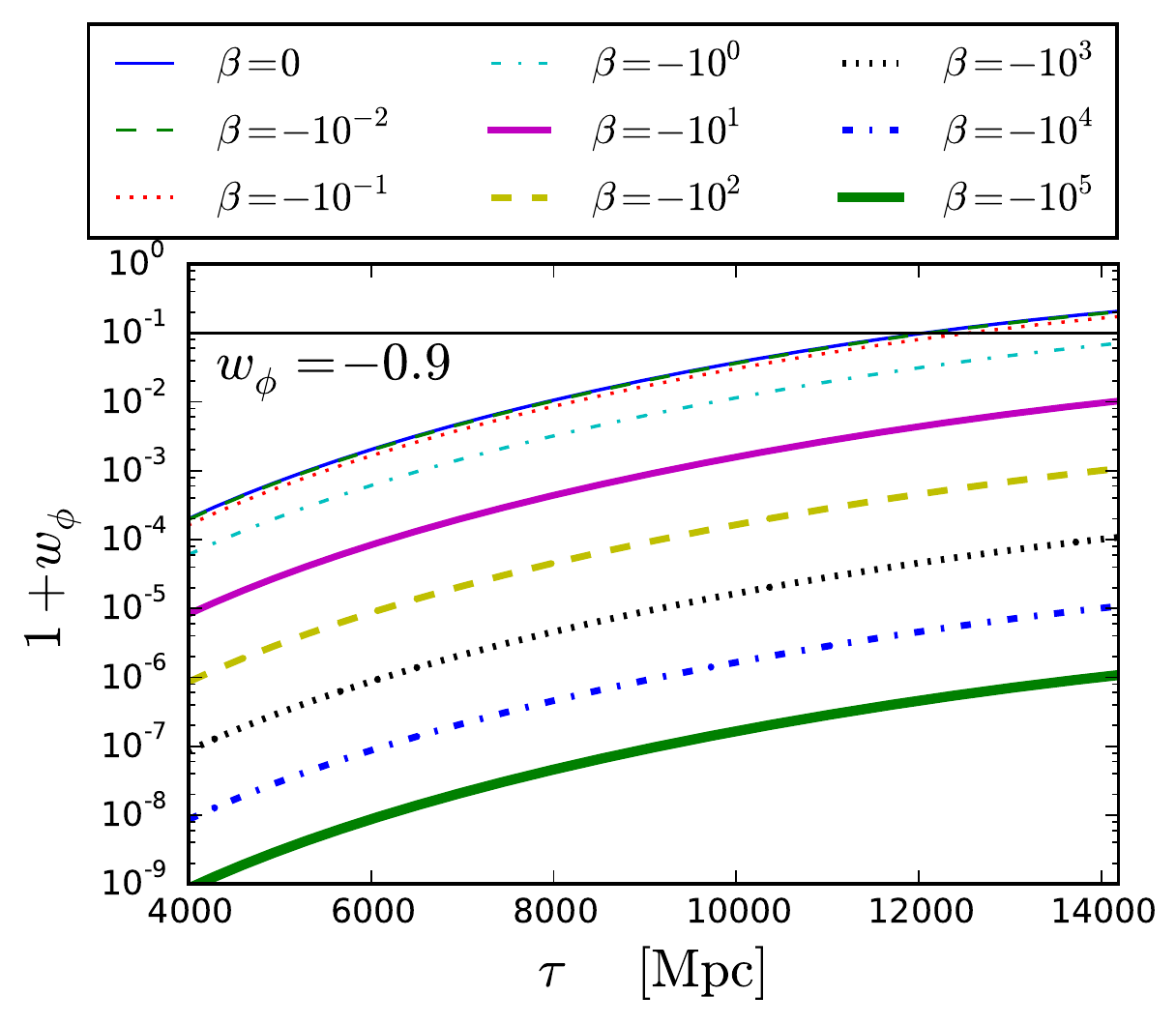}
\caption{ The evolution of the dark energy equation of state parameter $w_\phi = \Pb_\phi /\rhob_\phi$ as a function of the coupling parameter $\beta$. A constant $w_\phi = -0.9$ is shown for comparison. }
\label{fig:wbeta}
\end{figure}

For completeness, we also show the evolution of the dark energy equation of state parameter $w_\phi = p_\phi /\rho_\phi$ in Fig.~\ref{fig:wbeta}. We see that as the magnitude of the $\beta$ parameter increases, $w_\phi$ gets closer to $-1$, i.e. the background dark energy evolution resembles that of a cosmological constant. That is because of the effects of the coupling on the evolution of $\phib$: the $\beta \dot{\phib}^2/a^2$ term becomes completely subdominant to $V(\phi)$ and $w_\phi \rightarrow -1$.

Before we move on to our full MCMC analysis, we should stress that the effect of growth suppression in coupled dark energy models is quite rare and difficult to achieve. 
The same is true for modified gravity models, the vast majority of which exhibit growth increase \footnote{This is easy to understand as such models result to an enhanced effective gravitational constant $G_{\rm eff}$ with respect to Newton's constant. However, a recent study of theories beyond Hordenski showed that there is the possibility of constructing stable and ghost-free weak gravity scenarios~\cite{Tsujikawa:2015mga}.
} and make the tension worse as they favour a large $\sigma_8$ value~\cite{Hu:2015rva}.
In coupled dark energy models that exhibit both energy and momentum exchange, like the coupled quintessence model with $Q=\alpha_0 \dot{\bar{\phi}}\rhob_c$~\cite{Amendola:1999er}, the growth rate depends on two terms: a fifth-force contribution $\propto \alpha^2_0$, and a friction term $\propto \alpha_0 \dot{\phib}$, whose sign is determined by the sign of the coupling parameter $\alpha_0$ and the quintessence potential through the $\dot{\phib}$ dependence. 
In general, getting suppression of growth is highly non-trivial, and often requires potentials with more than one free parameters in order to make the friction term dominate over the competing fifth-force term, which tends to give growth increase for positive and negative coupling values (see~\cite{Tarrant:2011qe} for details and further discussion). 

On the contrary, our pure momentum transfer model has a straightforward behaviour for the simplest case of the 1EXP potential: Considering the positive coupling Type 3 model and the negative coupling T3 model we 
get growth increase for the former case and suppression of growth for the latter. 
This effect comes from the modified density contrast evolution due to the presence of the momentum transfer coupling, and there is no competing coupling term in Equation~(\ref{eq:deltac}). 
For the phenomenological $\text{T3}_\text{ph}$ model we also get growth suppression for a very wide range of negative $\beta$, but due to the scale dependent feature demonstrated in Fig.~\ref{fig:ClTT_T3ph} we also have a turning point where growth and $\sigma_8$ start to increase. For these reasons we chose to perform our MCMC analysis on the T3 and $\text{T3}_\text{ph}$ models separately.

\section{Results}
\label{sec:results}

\subsection{Datasets and priors}

\begin{figure*}[tb]
\centering
\includegraphics[width=0.495\linewidth]{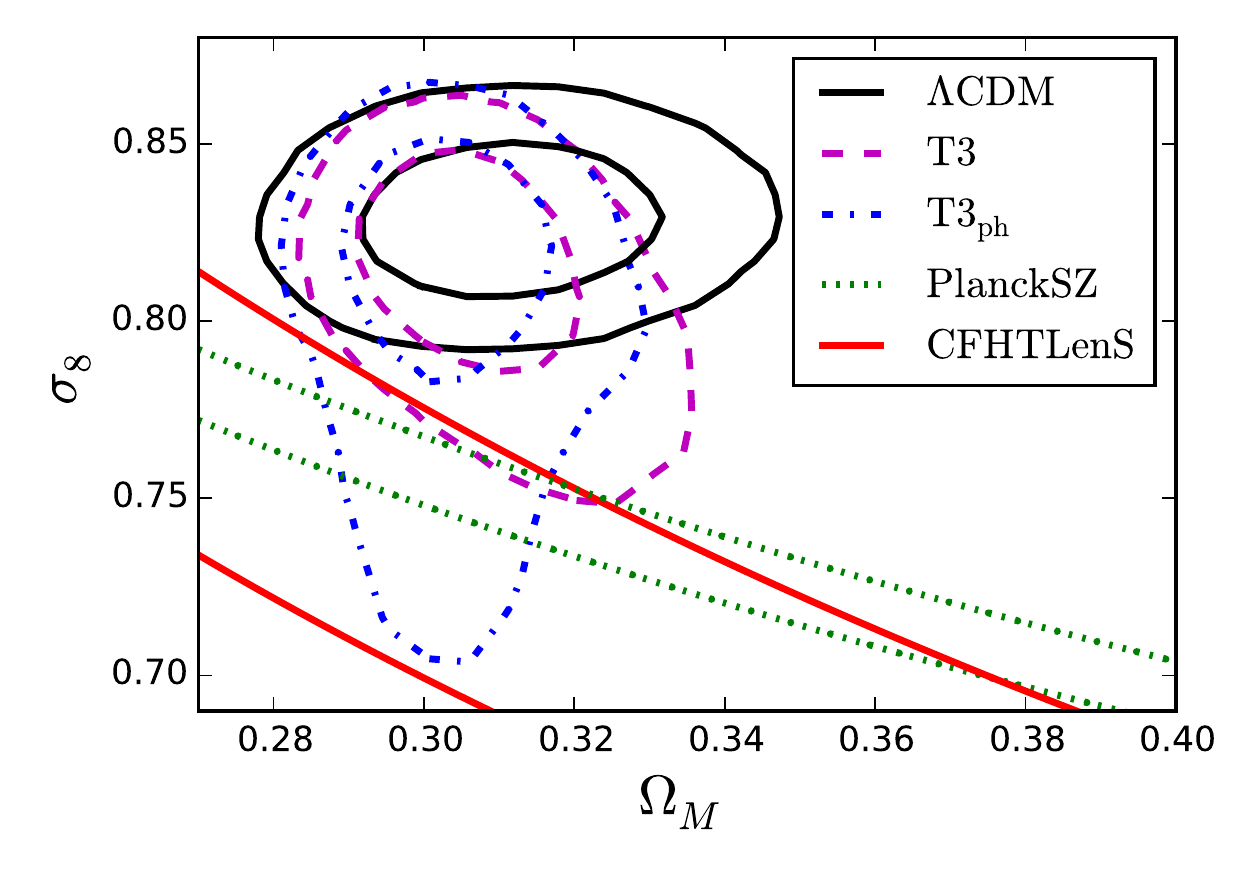}
\hfill
\includegraphics[width=0.495\linewidth]{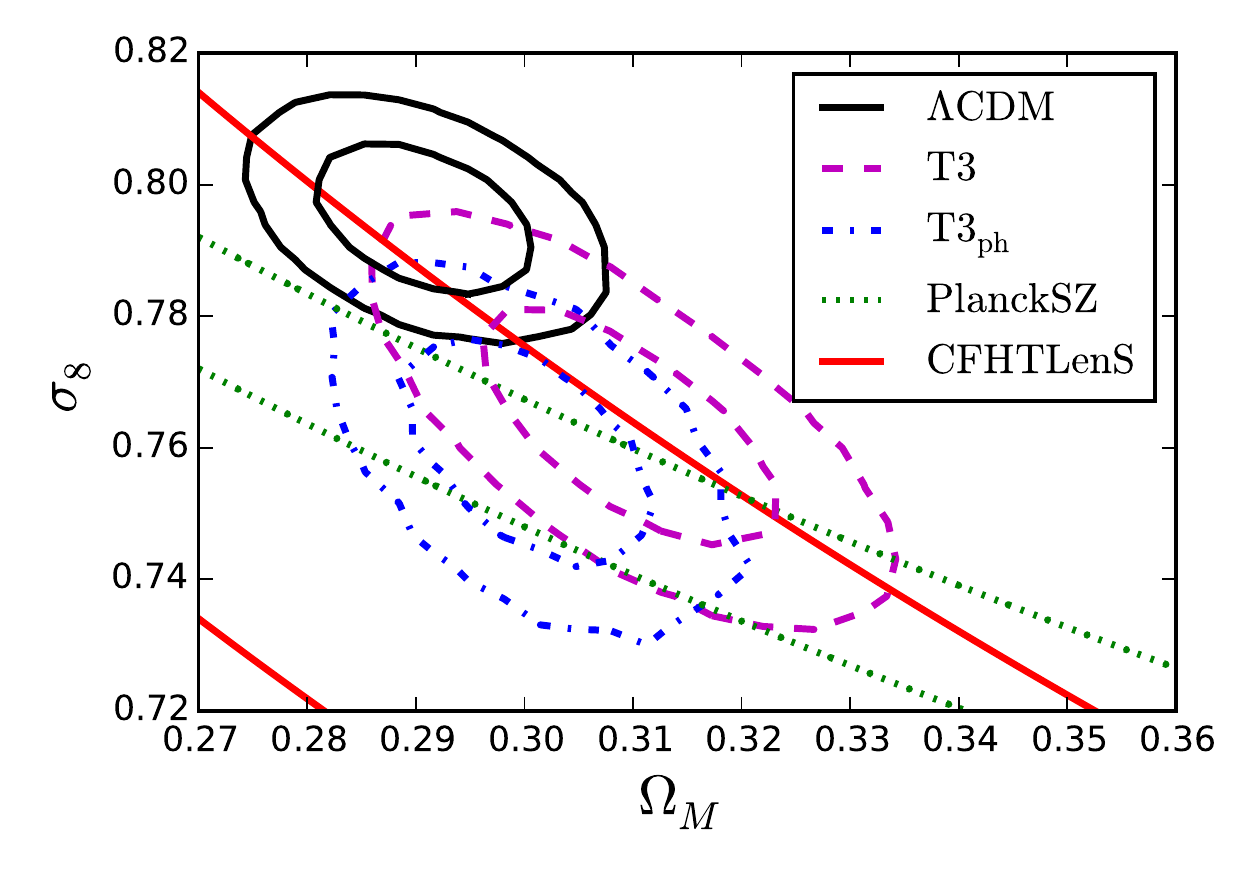}
\caption{$1\sigma$ and $2\sigma$ constraints for $\Lambda$CDM and the T3 and $\text{T3}_\text{ph}$ models in the $(\sigma_8,\Omega_M)$-plane using TT data (left panel) and using all data (right panel). The $1\sigma$-band from SZ clusters and CFHTLenS is shown in dotted green and solid red, respectively. $\Lambda$CDM is in tension with the SZ clusters and CFHTLenS while the T3 and $\text{T3}_\text{ph}$ models are compatible. Even after including the SZ-data, the $2\sigma$ $\Lambda$CDM contour does not overlap with the $1\sigma$ SZ contour, which illustrates the tension. On the contrary, the T3 and $\text{T3}_\text{ph}$ contours overlap with the SZ contour.}
\label{fig:plot3}
\end{figure*}

In our analysis we use a variety of recent datasets, including CMB data, baryon acoustic oscillation measurements, supernovae, and cluster counts. The specific likelihoods we employ are:
\begin{description}[leftmargin=4em,style=nextline]
\item[TT:] $C_\ell^{TT}$-data from Planck 2015~\cite{Ade:2015xua}, including low-$\ell$ polarisation.
\item[CMB:] TT and the lensing reconstruction from Planck 2015 data~\cite{Ade:2013tyw}.
\item[B:] Baryon Acoustic Oscillation (BAO) data from BOSS~\cite{Anderson:2013zyy}.
\item[J:] Joint Light-curve Analysis (JLA)~\cite{Betoule:2014frx}.
\item[SZ:] Planck SZ cluster counts~\cite{Ade:2013lmv,Ade:2015fva}.
\end{description}
We did not include weak lensing data in the MCMC run --- however, we show the constraint $\sigma_8\left(\Omega_M/0.27 \right)^{0.46} = 0.774 \pm0.040$ derived from CFHTLenS~\cite{Heymans:2013fya} in Fig.~\ref{fig:plot3}. Note that this constraint has been inferred assuming the $\Lambda$CDM model. 

We must also comment on the applicability of the cluster count likelihood we use~\cite{Ade:2013lmv}. This likelihood assumes a mass bias $(1-b)\simeq 0.8$ that agrees well with simulations, but the authors note that a significantly lower value would alleviate the CMB-SZ tension. It also uses the Tinker et al halo mass function~\cite{Tinker:2008ff}, which has been calibrated against N-body simulations assuming $\Lambda$CDM. This implies that including this likelihood is not entirely self-consistent, see for instance Ref.~\cite{Cataneo:2014kaa} for a discussion of the similar problem in $f(R)$ gravity. Making quantitative predictions for the non-linear effects of our model is not possible at this stage. However, a very recent paper on structure formation simulations with momentum exchange showed that the qualitatively similar dark scattering model~\cite{Simpson:2010vh} can alleviate the CMB-LSS tensions while keeping non-linear effects very mild~\cite{Baldi:2016zom}. Running a suite of N-body simulations for a coupled quintessence model with pure momentum exchange is the subject of future work.

We chose flat priors on the following set of cosmological parameters,
\begin{equation}
\left\{\omega_b, \omega_\text{cdm}, \theta_s, A_s, n_s, \tau_\text{reio}, \lambda\right\},
\end{equation}
and the collection of nuisance parameters required by the Planck and JLA likelihoods. The prior ranges of $\lambda$ and $\beta$ were chosen as
\begin{align}
\lambda &\in [0; 2.1], & \beta &\in [-0.5,0],
\intertext{for the T3 model with a flat prior on $\beta$ and }
\lambda &\in [0; 2.1], & {\rm log}_{10}(-\beta) &\in [-3,7],
\end{align}
for the $\text{T3}_\text{ph}$ model with a logarithmic prior on $\beta$ as indicated. As we have already mentioned, we exclude the $\beta>0$ branch from our analysis since we want to focus on the branch which can provide suppression of growth.
The initial conditions for the quintessence field were chosen as $(\phi_i, \dot{\phi}_i)=(10^{-4},0)$.

\begin{figure}[tb]
\centering
\includegraphics[width=\columnwidth]{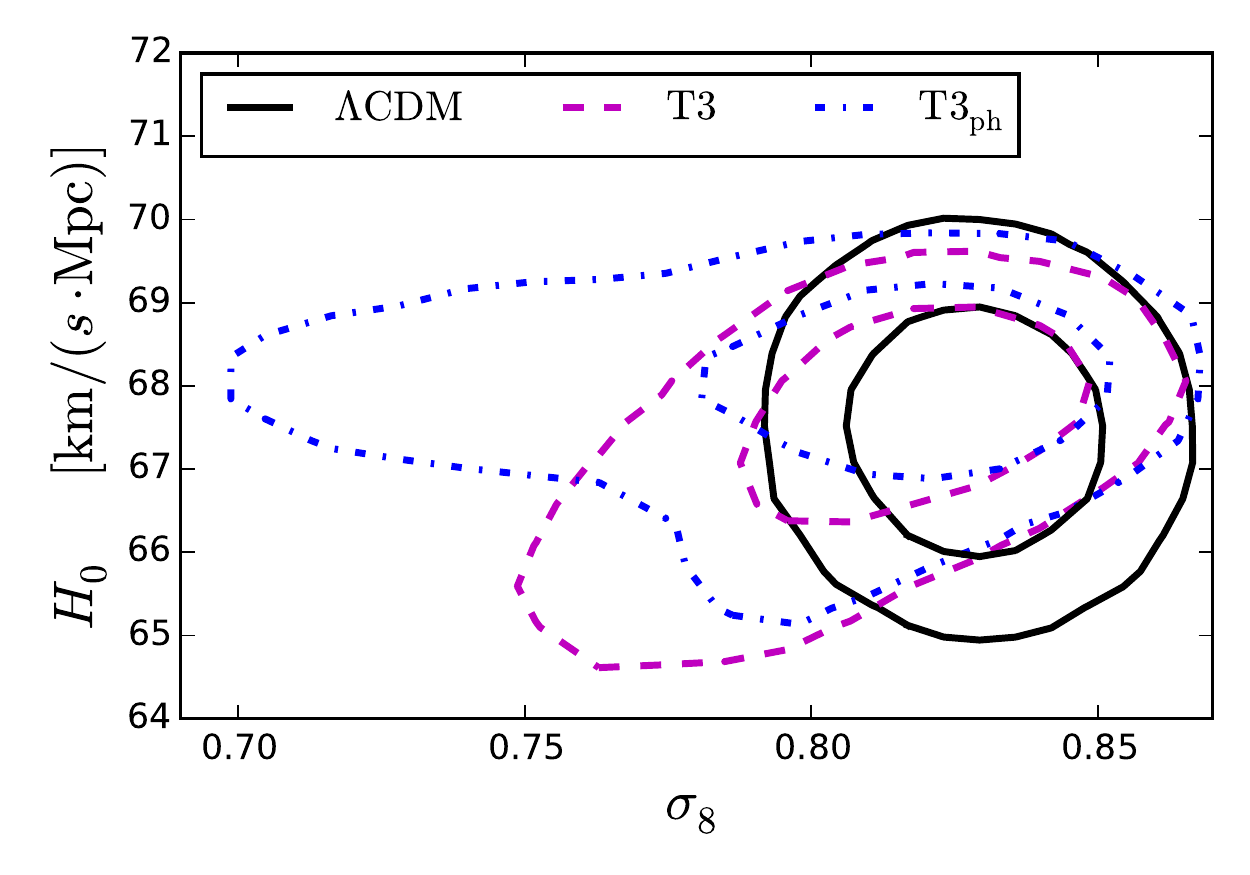}
\caption{$1\sigma$ and $2\sigma$ temperature (TT) constraints in the $(\sigma_8,H_0)$-plane for $\Lambda$CDM and the T3 and $\text{T3}_\text{ph}$ models.  The two parameters are uncorrelated in $\Lambda$CDM and $\text{T3}_\text{ph}$ while they are correlated in the T3 model}.
\label{fig:plot2b}
\end{figure}

\subsection{Parameter inference}

We perform a Markov chain Monte Carlo (MCMC) analysis using the publicly available code~\MontePython{}.
In the left panel of \fref{fig:plot3} we show the constraint from TT data alone in the $(\Omega_M,\sigma_8)$-plane for $\Lambda$CDM and the T3 and $\text{T3}_\text{ph}$ models. The $2\sigma$-contours of the T3 and $\text{T3}_\text{ph}$ models overlap with the $1\sigma$-constraint from SZ clusters  and CFHTLenS whereas the $\Lambda$CDM model is in tension with both SZ and CFHTLenS data.

\begin{figure*}[tb]
\centering
\includegraphics[width=0.495\linewidth]{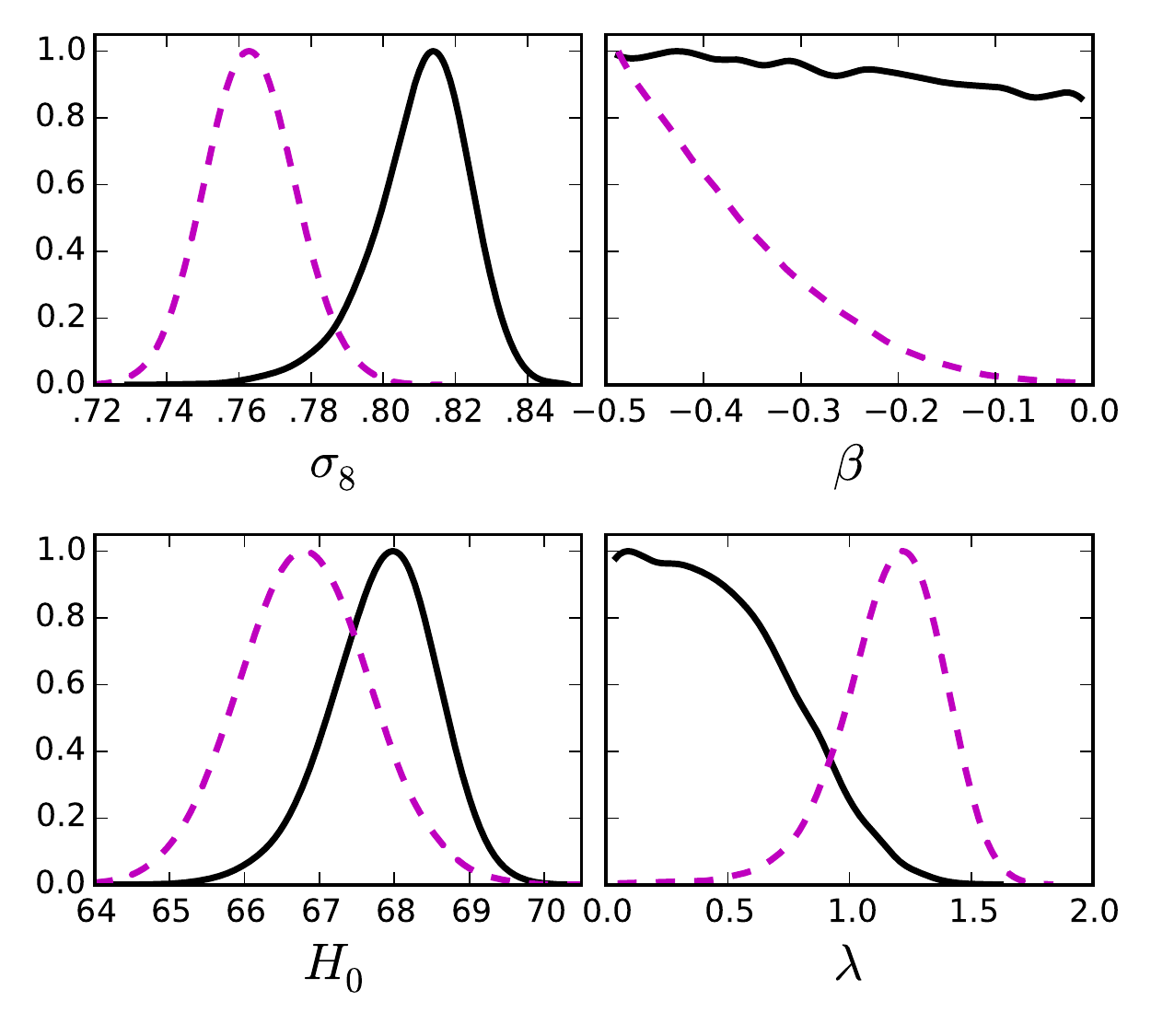}
\hfill
\includegraphics[width=0.495\linewidth]{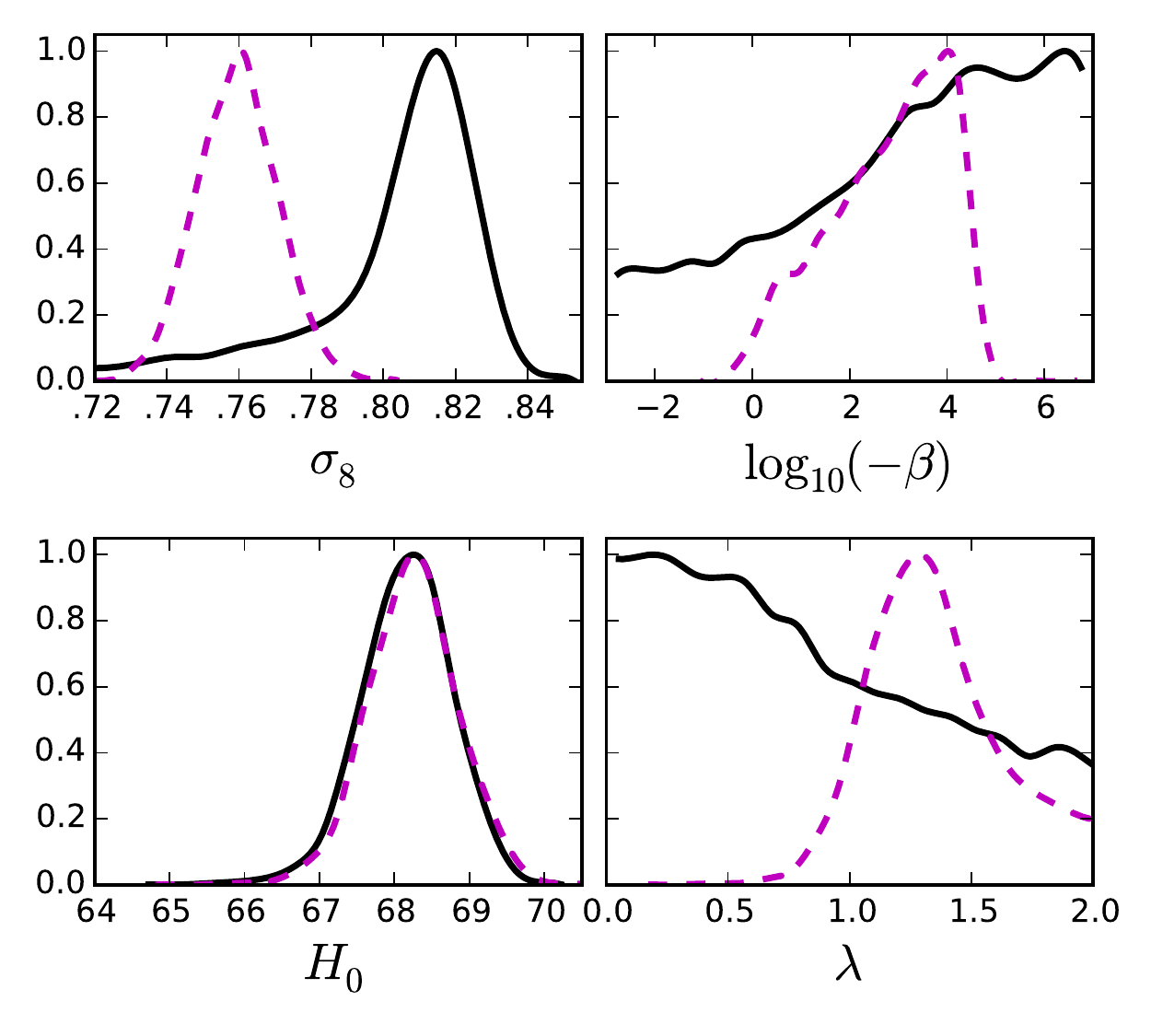}
\caption{One dimensional posterior distributions of the parameters $\{\sigma_8, H_0, \beta, \lambda\}$ excluding (solid black lines) and including (magenta dashed lines) the SZ cluster data along with the rest of our datasets. In the left panel we show the T3 model while the right panel shows the $\text{T3}_\text{ph}$ model.}
\label{fig:plot4}
\end{figure*}

When we combine all our datasets including the SZ cluster data we find quite different $(\Omega_M,\sigma_8)$-constraints as illustrated in the right panel of \fref{fig:plot3}.  The T3 and $\text{T3}_\text{ph}$ models generally prefer lower $\sigma_8$ and larger $\Omega_M$-values compared to $\Lambda$CDM. As we shall see in \sref{sec:chisq} the T3 and $\text{T3}_\text{ph}$ models are significantly better fits to the data than $\Lambda$CDM. In \fref{fig:plot2b} we show the $(\sigma_8,H_0)$-constraints. In the T3 model, $\sigma_8$ and $H_0$ become strongly correlated, but in $\Lambda$CDM and $\text{T3}_\text{ph}$ no correlation exists. As we will see later, this difference between T3 and $\text{T3}_\text{ph}$ has important implications for the $H_0$ tension.

\begin{figure}[tb]
\centering
\includegraphics[width=\columnwidth]{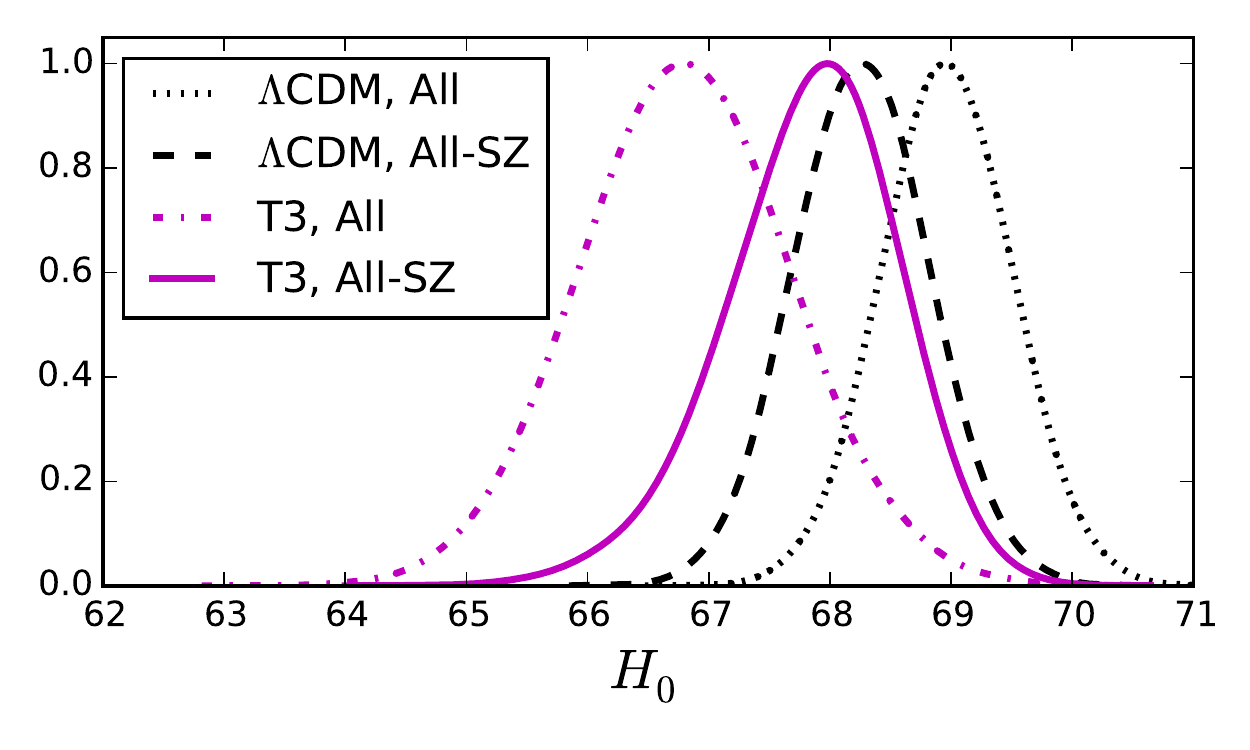}
\caption{One dimensional posterior distributions of the Hubble parameter. The inclusion of SZ clusters drives $H_0$ to larger values in $\Lambda$CDM and lower values in T3.}
\label{fig:plot5}
\end{figure}

In \fref{fig:plot4} we show the effect of including the SZ cluster data or not on the parameters $\{\sigma_8, H_0, \beta, \lambda\}$ for the T3 model (left panel) and $\text{T3}_\text{ph}$ model (right panel). Let us first comment on the T3 case: Taken at face value, the cluster data rules out the non-interacting case, $\beta=0$. Looking at the posterior distribution for $\lambda$ reveals that the non-cluster datasets roughly prefer $\lambda < 1$ while we have $\lambda > 1$ after the inclusion of the cluster data. This suggests that another potential could give an even better fit to the data than the single exponential potential  (but would probably rely on additional free parameters). The T3 posterior distribution for $\beta$ suggests that a better fit can possibly be found if we let $\beta < -1/2$. Moving on to the $\text{T3}_\text{ph}$ case we verify this as the posterior distribution peaks at $\beta=-10^4$. 
We also note that the $H_0$ distributions for $\Lambda$CDM and $\text{T3}_\text{ph}$ are very similar. In $\Lambda$CDM it is well known that the inclusion of SZ cluster data drives $H_0$ to larger values. In the T3 model the opposite happens, the cluster data pushes $H_0$ to lower values. We illustrate these effects in \fref{fig:plot5}.
This does put the T3 model in mild tension with local measurements of the Hubble constant but not much more than pure $\Lambda$CDM. However, as we already saw in the $\text{T3}_\text{ph}$ model the $H_0$ posterior distribution is very similar to the $\Lambda$CDM one.

In Table~I we show the mean values and $1\sigma$ confidence intervals on the cosmological parameters for various datasets combinations for the $\Lambda$CDM, T3, and $\text{T3}_\text{ph}$ models. It is important to note that $\beta$ is essentially not constrained unless we include the SZ cluster data in the analysis.
\begin{table*}
\caption{\label{tab:parameters}Cosmological parameters for $\Lambda$CDM and the T3-model for the 4 different data set combinations used in the plots. Note that $\beta$ is essentially unconstrained unless cluster data is added.}
\begin{ruledtabular}
\begin{tabular}{l|rrr|rrr|rrr|}
 & \multicolumn{3}{c|}{CMB} & \multicolumn{3}{c|}{CMB+B+J} & \multicolumn{3}{c|}{CMB+B+SZ+J}\\
 & $\Lambda$CDM & T3 & $\text{T3}_\text{ph}$ & $\Lambda$CDM & T3 & $\text{T3}_\text{ph}$ & $\Lambda$CDM & T3 & $\text{T3}_\text{ph}$\\ \hline 
$100~\omega_{b }$ & $2.22_{-0.02}^{+0.02}$ & $2.22_{-0.02}^{+0.02}$ & $2.22_{-0.02}^{+0.02}$ & $2.23_{-0.02}^{+0.02}$ & $2.23_{-0.02}^{+0.02}$ & $2.23_{-0.02}^{+0.02}$ & $2.24_{-0.02}^{+0.02}$ & $2.25_{-0.02}^{+0.02}$ & $2.23_{-0.02}^{+0.02}$\\
$\omega_\text{cdm}$ & $0.119_{-0.002}^{+0.002}$ & $0.119_{-0.002}^{+0.002}$ & $0.119_{-0.002}^{+0.002}$ & $0.118_{-0.001}^{+0.001}$ & $0.117_{-0.001}^{+0.001}$ & $0.118_{-0.001}^{+0.001}$ & $0.116_{-0.001}^{+0.001}$ & $0.116_{-0.001}^{+0.001}$ & $0.118_{-0.001}^{+0.001}$\\
$10^4\theta_s$ & $104.20_{-0.04}^{+0.04}$ & $104.20_{-0.04}^{+0.04}$ & $104.20_{-0.04}^{+0.04}$ & $104.21_{-0.04}^{+0.04}$ & $104.21_{-0.04}^{+0.04}$ & $104.21_{-0.04}^{+0.04}$ & $104.20_{-0.04}^{+0.04}$ & $104.22_{-0.04}^{+0.04}$ & $104.21_{-0.04}^{+0.04}$\\
$10^9 A_s$ & $2.15_{-0.07}^{+0.06}$ & $2.16_{-0.07}^{+0.06}$ & $2.16_{-0.07}^{+0.06}$ & $2.18_{-0.06}^{+0.05}$ & $2.19_{-0.06}^{+0.05}$ & $2.18_{-0.06}^{+0.05}$ & $2.08_{-0.05}^{+0.05}$ & $2.18_{-0.06}^{+0.06}$ & $2.19_{-0.06}^{+0.05}$\\
$n_{s }$ & $0.967_{-0.006}^{+0.006}$ & $0.967_{-0.006}^{+0.006}$ & $0.967_{-0.006}^{+0.006}$ & $0.970_{-0.005}^{+0.005}$ & $0.970_{-0.005}^{+0.005}$ & $0.970_{-0.005}^{+0.005}$ & $0.971_{-0.005}^{+0.004}$ & $0.974_{-0.005}^{+0.005}$ & $0.970_{-0.005}^{+0.004}$\\
$\tau_\text{reio}$ & $0.07_{-0.02}^{+0.02}$ & $0.07_{-0.02}^{+0.02}$ & $0.07_{-0.02}^{+0.02}$ & $0.08_{-0.01}^{+0.01}$ & $0.08_{-0.01}^{+0.01}$ & $0.08_{-0.01}^{+0.01}$ & $0.06_{-0.01}^{+0.01}$ & $0.08_{-0.02}^{+0.02}$ & $0.08_{-0.01}^{+0.01}$\\
$\Omega_M$ & $0.31_{-0.01}^{+0.01}$ & $0.32_{-0.03}^{+0.01}$ & $0.31_{-0.02}^{+0.01}$ & $0.301_{-0.007}^{+0.007}$ & $0.304_{-0.009}^{+0.008}$ & $0.301_{-0.008}^{+0.007}$ & $0.291_{-0.007}^{+0.007}$ & $0.310_{-0.010}^{+0.010}$ & $0.300_{-0.008}^{+0.008}$\\
$\sigma_{8 }$ & $0.818_{-0.010}^{+0.010}$ & $0.79_{-0.01}^{+0.03}$ & $0.796_{-0.009}^{+0.034}$ & $0.819_{-0.009}^{+0.009}$ & $0.81_{-0.01}^{+0.02}$ & $0.801_{-0.006}^{+0.030}$ & $0.795_{-0.008}^{+0.008}$ & $0.76_{-0.01}^{+0.01}$ & $0.76_{-0.01}^{+0.01}$\\
$H_0$ & $67.8_{-1.0}^{+0.9}$ & $66.2_{-1.1}^{+2.4}$ & $67.1_{-0.8}^{+1.7}$ & $68.3_{-0.6}^{+0.6}$ & $67.8_{-0.6}^{+0.8}$ & $68.2_{-0.6}^{+0.6}$ & $69.0_{-0.6}^{+0.6}$ & $66.8_{-0.9}^{+0.9}$ & $68.2_{-0.6}^{+0.6}$\\
$\lambda$ & \multicolumn{1}{c}{---} & $0.8_{-0.8}^{+0.2}$ & $0.9_{-0.9}^{+0.3}$ & \multicolumn{1}{c}{---} & $0.5_{-0.5}^{+0.1}$ & $0.8_{-0.8}^{+0.3}$ & \multicolumn{1}{c}{---} & $1.2_{-0.2}^{+0.2}$ & $1.4_{-0.4}^{+0.2}$\\
$\beta$ & \multicolumn{1}{c}{---} & $-0.24_{-0.09}^{+0.24}$ & \multicolumn{1}{c|}{---} & \multicolumn{1}{c}{---} & $-0.26_{-0.24}^{+0.07}$ & \multicolumn{1}{c|}{---} & \multicolumn{1}{c}{---} & $-0.39_{-0.11}^{+0.03}$ & \multicolumn{1}{c|}{---}\\
$\log_{10}(-\beta)$ & \multicolumn{1}{c}{---} & \multicolumn{1}{c}{---} & $2.5_{-5.5}^{+4.5}$ & \multicolumn{1}{c}{---} & \multicolumn{1}{c}{---} & $3.1_{-1.1}^{+3.9}$ & \multicolumn{1}{c}{---} & \multicolumn{1}{c}{---} & $2.8_{-0.8}^{+1.7}$\\
\end{tabular}
\end{ruledtabular}
\end{table*}

\subsection{$\chi^2$-values}\label{sec:chisq}

 In Table~II we show the $\chi^2$ values for the best-fitting $\Lambda$CDM, T3 and $\text{T3}_\text{ph}$ models.  Our T3 and $\text{T3}_\text{ph}$ models can reconcile CMB, BAO and LSS data. As can be seen from the last two lines in Table~II, when the SZ cluster dataset is included the preference for the T3 and $\text{T3}_\text{ph}$ models is strong. 

\begin{table}
\caption{\label{tab:chisq}$\chi^2$-values for $\Lambda$CDM, the T3-model and the T3ph-model for all tested datasets. The preference for T3 and $\text{T3}_\text{ph}$ is strong when cluster data is included.}
\begin{ruledtabular}
\begin{tabular}{l|rrrrr}
Dataset & $\chi^2_{\Lambda\text{CDM}}$ & $\chi^2_\text{T3}$ & $\chi^2_{\text{T3}_\text{ph}}$ &$\Delta \chi^2_\text{T3}$ & $\Delta \chi^2_{\text{T3}_\text{ph}}$\\
\hline
TT & $11261.80$ & $11265.12$ & $11265.20$ & $-3.32$ & $-3.40$\\
TT+J & $11946.80$ & $11949.54$ & $11950.34$ & $-2.74$ & $-3.54$\\
CMB & $11271.80$ & $11271.78$ & $11273.18$ & $0.02$ & $-1.38$\\
CMB+J & $11956.52$ & $11956.86$ & $11957.26$ & $-0.34$ & $-0.74$\\
CMB+B & $11274.46$ & $11275.58$ & $11274.88$ & $-1.12$ & $-0.42$\\
CMB+B+J & $11958.80$ & $11958.68$ & $11960.22$ & $0.12$ & $-1.42$\\
CMB+B+SZ & $11293.50$ & $11279.44$ & $11276.30$ & $14.06$ & $17.20$\\
CMB+B+SZ+J & $11978.38$ & $11965.84$ & $11961.90$ & $12.54$ & $16.48$\\
\end{tabular}
\end{ruledtabular}
\end{table}

\section{Conclusions}
\label{sec:conclusions}

We have identified an interacting dark energy model which can suppress structure growth and can reconcile CMB and LSS observations. It is a pure momentum transfer model and belongs to a class of theories constructed using the Lagrangian pull-back formalism for fluids --- the coupling function characterising the theory is not added at the level of the equations, but at the level of the action. In this way various pathologies and instabilities can be very easily identified; considering the ghost-free branch of the model, we investigated its observational signatures on the CMB and linear matter power spectra. For a constant coupling parameter $\beta$ and our specific choice of potential, the model exhibits structure growth for positive $\beta$ and growth suppression for negative $\beta$. Focusing on the latter case, we performed an MCMC analysis and found that using CMB and BAO data our model is as good as $\Lambda$CDM, while adding cluster data our model becomes strongly prefered. We note, however, that it still exhibits tension with local measurements of the Hubble constant. A full model selection analysis based on the Bayesian evidence is left for future work. However, we note that when the likelihood method shows preference for an extra parameter at a level of $3 \sigma$, so does the Bayesian analysis which is also quite sensitive to the priors used (see~\cite{Battye:2014qga} for a related analysis and discussion for the case of massive neutrinos). 

Our work offers a promising alternative for resolving the CMB and LSS tension. 
Another alternative is massive neutrinos, which have also been proposed to lift the discrepancy~\cite{Battye:2013xqa} but they increase the tension with the Hubble constant~\cite{Giusarma:2013pmn}. A recent interesting proposal was presented in~\cite{Lesgourgues:2015wza}, in which dark matter interacts with a new form of dark radiation and structure growth is damped via momentum transfer effects.
On the other hand, there is also the possibility that this tension is a result of poorly understood systematic effects. In the imminent future, a set of larger and better optical large scale structure surveys (the Dark Energy Survey, the Euclid satellite, the Large Synoptic Survey Telescope) as well as new probes with completely different methodology and systematics (e.g. 21cm intensity mapping with the Square Kilometre Array~\cite{Santos:2015gra}) will either resolve this tension or confirm the exciting prospect of new physics. 

In order to take full advantage of current and future large scale structure datasets, understanding of the non-linear effects of exotic dark energy models is crucial. For example, in order to use the full range of the available data with confidence, one needs to correct the power spectrum on small (non-linear) scales. $N$-body simulations related to pure momentum transfer in the dark sector have been performed in~\cite{Baldi:2014ica, Baldi:2016zom}, based on the elastic scattering model presented in~\cite{Simpson:2010vh}. We plan to investigate non-linear effects for the negative coupling Type-3 models in future work.

To conclude, we may have discovered a whole family of models (Type-3-like models with pure momentum transfer) that can give suppression of growth and reconcile the tension between CMB and LSS. In this work we focused on a particular case, but there is a plethora of different choices that give models belonging to the same class. For example, we could use another coupling function $h(Z)$ and/or another form for the quintessence potential. However, this way the results are strongly model-dependent. Using the PPF approach developed in~\cite{Skordis:2015yra}, we can try to parametrise the free non-zero functions that define Type-3 models in a model-independent way and study their observational consequences; arguably, we should be able to constrain these free functions (or their combinations) such that they give late time growth suppression relative to $\Lambda$CDM --- which is what the available data currently prefer. This would be very important for phenomenological model building and for determining the constraining and discriminating power of future surveys.

\FloatBarrier 
\section{Acknowledgments}
A.P. acknowledges support by STFC grant ST/H002774/1. T.T. acknowledges support by STFC grant ST/K00090X/1. Numerical computations were performed using the Sciama High Performance Compute (HPC) cluster which is supported by the ICG, SEPNet and the University of Portsmouth. We are indebted to Robert Crittenden for very useful comments and feedback.
We would also like to thank Ed Copeland, Kazuya Koyama, Jeremy Sakstein and Constantinos Skordis for useful discussions.

\FloatBarrier
\bibliographystyle{apsrev}
\bibliography{references}

\end{document}